 \documentclass[12pt,preprint]{aastex}

\slugcomment{}

\shorttitle{X-ray Flares in the Plunging Region}
\shortauthors{M. Machida & R. Matsumoto}

\begin{document}

\title{Global Three-Dimensional MHD Simulations of 
Black Hole Accretion Disks: 
X-ray Flares in the Plunging Region}

\author{Mami Machida}
\affil{Graduate School of Science and Technology, Chiba University, 
 1-33 Yayoi-cho, Inage-ku, Chiba 263-8522, Japan}
\email{machida@astro.s.chiba-u.ac.jp}

\and

\author{Ryoji Matsumoto}
\affil{Department of Physics, Faculty of Science, Chiba University, 
1-33 Yayoi-cho, Inage-ku, Chiba 263-8522, Japan}

\begin{abstract}
We present the results of three-dimensional global resistive 
magnetohydrodynamic (MHD) simulations
of black hole accretion flows. General relativistic effects are simulated 
by using the pseudo-Newtonian potential. 
Initial state is an equilibrium model of 
a torus threaded by weak toroidal magnetic fields. 
As the magnetorotational instability (MRI) grows in the torus, 
mass accretes to the black hole 
by losing the angular momentum.
We found that 
in the innermost plunging region,  
non-axisymmetric accretion flow creates bisymmetric spiral 
magnetic fields and current sheets.
Mass accretion along the spiral channel creates 
one armed spiral 
density distribution. 
Since the accreting matter carries in magnetic fields which subsequently 
are stretched and amplified due to differential rotation, 
current density increases inside the channel. 
Magnetic reconnection taking place in the current sheet 
produces slow mode shock waves which propagate away from the 
reconnection site. 
Magnetic energy release in the innermost plunging region can be the 
origin of X-ray shots observed in black hole candidates. 
Numerical simulations reproduced soft X-ray excess preceding 
the peak of the shots, X-ray hardening at the peak of the shot, 
and hard X-ray time lags. 
\end{abstract}


\section{Introduction}
Magnetic fields
play essential roles in various activities associated with accretion disks. 
The most important role is the angular momentum transport 
which enables the accretion of the disk material. 
Since Balbus \& Hawley (1991) pointed out the importance of the 
magneto-rotational instability (MRI) 
in accretion disks, the nonlinear growth 
of the instability has been studied by local three-dimensional 
magnetohydrodynamic (MHD) simulations (e.g., Hawley, Gammie, \& Balbus 
1995; Matsumoto \& Tajima 1995; Brandenburg et al. 1995)
and global MHD simulations (e.g., Armitage 1998; Matsumoto 1999; 
Hawley 2000; Machida, Hayashi, \& Matsumoto 2000). 

When general relativistic effects are taken into account, 
accretion flow in the innermost region of black hole accretion disks 
changes from a circularly rotating flow to a radially infalling flow. 
There exist some radius below 
which almost all energy is advected inward into the black hole.
The radial flow is transonic around the last stable orbit.
Steady models of black hole accretion disks including this 
transonic region were constructed under the prescription of 
$\alpha$-viscosity for optically thick disks 
(Muchotrzeb \& Paczy\'nski 1983; Matsumoto et al. 1984; 
Abramowicz et al. 1988) and for optically thin disks 
(e.g., Matsumoto, Kato \& Fukue 1985; 
Narayan, Kato \& Honma 1997). 
Global three-dimensional MHD simulations of black hole accretion flows 
have been carried out by Hawley (2000), Hawley (2001), 
Hawley \& Krolik (2001), Haley, Balbus \& Stone (2001), 
Armitage, Reynolds \& Chiang (2001), Reynolds \& Armitage (2001), 
Hawley \& Krolik (2002), Hawley \& Balbus (2002), 
and Krolik \& Hawley (2002) 
by using the pseudo-Newtonian potential 
(Paczy\'nski \& Witta 1980). 
In black hole accretion flows, 
they reproduced radial flow profiles similar to those obtained by 
assuming the phenomenological $\alpha$-viscosity. 
Hawley \& Krolik (2001) showed that the ratio of stress to pressure which 
corresponds to the $\alpha$ parameter in conventional disk models, 
exhibits both systematic gradients and large fluctuations; 
it rises from $10^{-2}$ in the disk midplane at large radius to 
$\sim 10$ near the midplane well inside the marginally stable radius. 
Thus the efficiency of angular momentum extraction in the plunging 
region can be larger than that in transonic disk models 
(see also Reylonds \& Armitage 2001). 

Another important mechanism in the innermost region of black hole 
accretion flow is the magnetic reconnection 
which can release magnetic energy even inside the marginally 
stable radius.
In differentially rotating disks, magnetic fields are stretched and 
amplified due to differential rotation. 
When uniform horizontal field threads an accretion disk, 
for example, 
magnetic fields twisted by differential rotation 
create current sheets which subject to magnetic reconnection. 
Tajima \& Gilden (1987) carried out MHD simulations of the formation
of current sheets and magnetic reconnection in accretion disks. 
They applied the mechanism 
to quasi-periodic oscillations of dwarf nova. 
Sano \& Inutsuka (2001) carried out three-dimensional 
local MHD simulations of MRI and showed quasi-periodic 
release of magnetic energy in accretion disks. 

In our previous papers (Machida et al. 2000; 
Kawaguchi et al. 2000; Machida et al. 2001), 
we showed the developments of turbulent magnetic fields and current 
sheets in accretion disks by three-dimensional global MHD simulations 
assuming ideal MHD. 
By evaluating the released magnetic energy 
from the current distribution obtained by ideal MHD simulations, 
Kawaguchi et al. (2000) successfully 
reproduced $1/f^{\alpha}$ type time variations observed in black hole 
candidates.

In addition to the fractal-like $1/f^{\alpha}$ time variations, 
Negoro et al. (1995) pointed out that the X-ray time 
variations from Cyg X-1 show large amplitude X-ray shots. 
The X-ray shots have typical time 
interval around several seconds and show symmetric time profile 
(Negoro et al. 1995; Negoro, Kitamoto, \& Mineshige 2001; 
Negoro \& Mineshige 2002). 
Manmoto et al. (1996) showed by hydrodynamical simulations based on 
$\alpha$-prescription of viscosity that inward propagating 
waves in advection dominated black hole accretion 
flows are reflected around the marginally stable radius  
and reproduced time symmetric X-ray profiles. 
They attributed the origin of rapid ($\sim$ ms) spectral hardening at 
the peak of X-ray shot to the temperature rise due to acoustic 
wave reflection. 

In this paper, we explicitly include the electric resistivity in 
basic equations and present the results of three-dimensional global 
resistive MHD simulations 
which show the magnetic energy release following the build up of 
current sheets in the 
innermost region of black hole accretion flows.

\section{Basic Equations and Simulation Model}
We solved the following resistive MHD equations in cylindrical 
coordinate system ($\varpi, \varphi, z$) 
by using a modified Lax-Wendroff scheme with artificial viscosity. 
We use the normalization $c = r_g = 1$ where $c$ is the light speed and 
$r_g$ is the Schwarzschild radius, 

\begin{equation}
   \frac{\partial \rho}{\partial t} 
    + \nabla (\rho \mbox{\boldmath $v$}) = 0
\end{equation}
%
%
\begin{equation}
   \rho \left[
   \frac{\partial \mbox{\boldmath $v$}}{\partial t}
    + \mbox{\boldmath $v$} \cdot \nabla \mbox{\boldmath $v$} 
        \right]
  =
   - \nabla P 
   + \mbox{\boldmath $j$}  \times \mbox{\boldmath $B$}
   - \rho \nabla \phi
\end{equation}
%
%
\begin{equation}
    \frac{\partial \mbox{\boldmath $B$}}{\partial t}
   =
    \nabla \times (\mbox{\boldmath $v$} \times \mbox{\boldmath $B$}
    - \eta \mbox{\boldmath $j$})
\end{equation}
%
%
\begin{equation}
   \rho T \frac{d S}{dt} 
      = 
    \eta j^2
\end{equation}
where $\rho$, $P$, $\mbox{\boldmath $v$}$, $\mbox{\boldmath $B$}$, 
$\phi$, $\mbox{\boldmath $j$}$, $T$ and $S$ are 
the density, pressure, velocity, magnetic fields, gravitational potential, 
current density, temperature and specific entropy, respectively. 
The specific entropy is expressed as 
$S = C_v \ln{(P/ \rho^{\gamma})}$ where $C_v$ is the specific heat at 
constant volume and $\gamma$ is the specific heat ratio. 
Joule heating term is included in the energy equation. 
We neglect radiative cooling. 
We assume the anomalous resistivity $\eta$ 
adopted in solar flare simulations (e.g., Yokoyama \& Shibata 1994) 
which mimics the enhancement of 
 resistivity 
where current density $j$ is large, as follows;
\begin{equation}
\eta = \eta_0 [ {\rm max}(v_d/v_c -1, 0)]^2
\end{equation}
where 
$v_d \equiv j / \rho$, and $v_c$ is the threshold above which 
anomalous resistivity sets in. 
We adopt $v_c = 8$, and 
$\eta_0 = 5 \times 10^{-4}$.

The initial condition is an equilibrium model of an axisymmetric MHD 
torus threaded by toroidal magnetic fields surrounding 
a black hole.
We assume pseudo-Newtonian potential $\phi = - GM/(r-r_g)$ 
(Paczy\'nski \& Witta 1980) where 
$G$ is the gravitational constant and 
 $M$ is the mass of the black hole.
We neglect the self-gravity of the gas. At the initial state, the torus 
is assumed to have a constant specific angular momentum $L$ and 
assumed to obey a polytropic equation of state 
$P=K\rho^{\gamma}$ where $K$ is a constant. 
We assume that the torus is embedded in hot, 
non-rotating, isothermal, spherical halo.

According to 
Okada, Fukue \& Matsumoto (1989),  
we assume the square of the Alf\'ven speed 
$v_{\rm A}^2 = B_{\varphi}^2/(4 \pi \rho) \equiv H(\rho \varpi^2)^{\gamma-1}$
, where $B_{\varphi}$ is the toroidal magnetic field and 
$H$ is a constant.  Using this assumption, we can integrate 
the equation of motion into a potential form;
\begin{equation}
\Psi(\varpi, z) = - \frac{1}{2(r - 1)} 
       + \frac{L^2}{2 \varpi^2}
       + \frac{1}{\gamma-1}v_s^2
       + \frac{\gamma}{2(\gamma-1)}v_{\rm A}^2
     = \Psi_b = constant  , 
\end{equation}
where $v_s^2 = \gamma P / \rho$ is the square of the sound speed, 
$r = ( \varpi^2 + z^2 )^{1/2}$, 
and $\Psi_b = \Psi(\varpi_b,0)$. 
Here, the reference radius $\varpi_b$ is defined as the radius where 
the rotation 
speed $L/\varpi_b$ equals 
the Keplerian velocity $v_{\rm Kb} = [\varpi_b
(\partial \phi / \partial \varpi)_{\varpi = \varpi_b}]^{1/2} $. 
By using equation (6), we obtain the density 
distribution as 
\begin{equation}
\rho = \rho_b \left\{
      \frac{ {\rm max}[\Psi_b + 1/(2(r-1)) - L^2/(2 \varpi^2),0]}
      {K [ \gamma/(\gamma-1)]
       [ 1 + \beta_b^{-1} \varpi^{2(\gamma-1)}/\varpi_b^{2(\gamma-1)}]}
       \right \}^{1/(\gamma-1)}
\end{equation} 
where $\beta_b \equiv (2K/H)/\varpi_b^{2(\gamma-1)}$ 
is the ratio of gas pressure to magnetic pressure at 
$(\varpi,z) = (\varpi_b,0)$.
We set $\rho_b = 1$. 
The parameters describing the structure 
of the MHD torus are $\gamma$, $\beta_b$, $L$, and $K$. In this paper 
we report the results of simulations for parameters 
$\varpi_b = 50 r_g$, $\beta_b = 100$, 
$\gamma = 5/3$, $ L = (\varpi_b/2)^{1/2} \cdot \varpi_b/(\varpi_b-1)$, 
and $K=0.0005$. The density and sound speed 
of the halo at $r = \varpi_b$ is taken to be 
$\rho_{\rm halo}/\rho_b = 10^{-3}$ 
and $v_{s,\rm{halo}} = 3/5$, respectively.

The number of mesh points  are $(N_{\varpi},N_{\varphi},N_z) =
(250, 64, 192)$. 
The grid size is $\Delta \varpi = 
\Delta z = 0.1$ for $ 0 < \varpi, z < 10 $, and otherwise 
increases with $\varpi$ and $z$. 
The outer boundaries at $\varpi = 150$
and at $z = 70$ are free boundaries where waves can transmit. 
A periodic boundary condition 
is imposed for the $\varphi$-direction. 
Since we include full azimuthal angle $(0 \leq \varphi \leq 2 \pi)$ 
into the simulation region, we can follow the evolution of 
low azimuthal wave number non-axisymmetric structures. 
We imposed symmetric boundary 
condition at the equatorial plane. 
Absorbing boundary condition is applied at the inner boundary 
by introducing a damping rate
\begin{equation}
a_i = 0.1( 1.0 - \tanh{
     \frac{r - 2 + 5 \Delta \varpi}{2 \Delta \varpi} }). 
\end{equation}
Inside $r = 2$, the deviation of physical quantity $q$ from initial 
value $q_0$ is artificially reduced with damping rate $a_i$ as 
\begin{equation}
 q^{new} = q - a_i ( q - q_0). 
\end{equation}
This damping layer serves as the non-reflecting boundary 
which absorbs accreting mass and waves propagating 
inside $r = 2$. 
A small-amplitude, 
random perturbations are imposed at $t = 0$ for the azimuthal velocity.

\section{Numerical Results}
\subsection{Global Structure of Non-Radiative Accretion Disk}
Figure 1a and figure 1b show the isosurface of the density $(\rho = 0.4)$ 
at $t = 0$ (initial condition), and $t = 30590 \simeq 10 t_0$ 
where $t_0 = 2 \pi \varpi_b / v_{Kb} \simeq 3079$  
is the one orbital time at $\varpi_b$. Owing to the efficient angular 
momentum redistribution, the torus is deformed into a disk and 
matter accretes to the central part by losing the angular momentum. In the 
outer part of the torus, matter gets angular momentum and expands radially. 
The angular momentum transport is 
initially driven by the magneto-rotational instability.  
After the non-axisymmetric MRI grows, 
the disk region becomes turbulent. 

Figure 2a shows the azimuthally averaged density contours and 
the momentum vectors at $t=30590$. 
Circulating motions similar to 
those observed in ideal MHD simulations (e.g., 
Hawley \& Krolik 2001; Machida et al. 2001) 
appear near the equatorial plane ($\varpi \sim 12$). 
In the innermost region ($\varpi < 5 $) radial infall becomes significant. 
Near the surface layer around $\varpi \sim 7$, shock wave front is formed 
and higher angular momentum gas blows out into the disk corona. 
Figure 2b shows the azimuthally averaged density distribution at $t=0$, 
$t=21502$ and $t=30590$. 

Figure 3a and 3c show the equatorial density distribution. 
Figure 3b and 3d show magnetic field lines projected onto the 
equatorial plane. 
In the outer region (figure 3a, 3b), magnetic field lines are tightly wound 
and show turbulent structure. 
In the inner region (figure 3c, 3d), 
magnetic field lines are less turbulent and 
globally show bisymmetric spiral (BSS) shape. 
One armed spiral mode dominates in the 
density distribution. 
Figure 4 shows three-dimensional structure of magnetic field lines at 
$t=30590$ around the surface of the innermost region ($\varpi < 10$) of 
the disk. White curves show the magnetic field lines. Yellow plane shows the 
equatorial plane and blue surface is the isosurface of the density 
($\rho = 0.4$). 
Magnetic field lines have significant $z$-components and show helical 
structure near the surface of the disk. 
Matter swirls into the black hole along these field lines. 
Bisymmetric spiral magnetic fields are created 
owing to the infall of the disk material. 

\subsection{Time Evolution and Angular Momentum Redistribution}

The amplification of magnetic fields and its saturation occurs similarly 
to those reported by Hawley \& Krolik (2001, 2002). 
Figure 5a shows the time evolution of 
the ratio of the gas pressure to magnetic pressure 
$\beta  = \langle P \rangle / 
\langle B^2/8 \pi \rangle$ where the bracket means the volume average. 
The solid curve and the dashed curve show the average 
in the inner region ($ 4 \leq \varpi \leq 10$, $0 \leq z \leq 1$) 
and in the outer region ($20 \leq \varpi \leq 40$, $0 \leq z \leq 3$), 
respectively. 
The disk stays in a quasi-steady state with $\beta \sim 10$ for 
time scale much longer than the dynamical time scale. 
Figure 5b shows the time evolution of 
the ratio of the Maxwell stress to pressure 
$\alpha_B \equiv - \langle B_{\varpi} B_{\varphi}/4 \pi \rangle / 
 \langle P \rangle$
averaged in the inner region and in the outer region.
We found $ \alpha_B \simeq 0.1$ in the inner region 
and $\alpha_B \simeq 0.02$ in the outer region.
Figure 5c shows the time variation of the accretion rate at 
$\varpi =2.5$. Accretion rate increases with time even at the 
end of simulation because much mass is still stored in the original torus 
($\varpi_b \sim 50$). It is beyond the scope of this paper to 
continue simulation for time scale long enough for the black hole 
to accrete most of the mass of the initial torus. 

The Maxwell stress in figure 5b 
shows spikes similar to those observed in local 
simulations. Sano \& Inutsuka (2001) showed by local three-dimensional 
resistive MHD simulations that these spikes are created by current 
dissipation in channel flows which appear in the nonlinear stage of the 
MRI. 
Although the BSS magnetic fields which appear in our simulation are not 
exactly the same as channel flows in a disk initially threaded by 
vertical magnetic fields, 
the BSS fields have current layers where the magnetic field lines change 
their direction. Magnetic reconnection can take place in such current layers. 
We call the region as BSS channel.  

Figure 6 shows the distribution of density, pressure, radial velocity 
and specific angular momentum 
near the equatorial plane averaged azimuthally, 
vertically ($0 < z < 0.3$) and in time. 
Solid curves and dashed curves show average in $29000 < t < 31000$ and 
$28000 < t < 29000$, respectively. 
Around $\varpi \simeq 10$, 
The radial structure changes from a dense torus to an accreting disk 
(figure 6a). 
The equatorial density increases with time because the torus flattens. 
Figure 6b shows the averaged gas pressure 
and the magnetic pressure. 
In the innermost region ($2 \leq \varpi \leq 3$), 
magnetic pressure is about $30$\% of the gas pressure. 
Figure 6c shows the distribution of the radial velocity. 
In the inflowing region ($\varpi \leq 10$), 
the radial velocity is roughly proportional to $\varpi^{-3}$ 
and exceeds $0.1c$ around the radius of marginally stable 
orbit ($\varpi \sim 3$). 
The radial velocity in the outer region ($\varpi > 10$) is 
much smaller than the sound speed. 
Figure 6d shows the specific angular momentum averaged in the vertical 
direction, azimuthal direction and in time. 
The dotted curve shows the Keplerian angular momentum distribution. 
Although the distribution of initial angular momentum is uniform, 
the angular momentum distribution becomes nearly Keplerian 
in the radial range $3 < \varpi < 100$. 

%
Figure 7 shows the time development of the density averaged 
azimuthally and vertically 
($0 \leq z \leq 0.3$). 
Gray scale shows the logarithmic density whose range is 
$-0.5$ (blue) $ \leq \log{\rho} \leq 0$ (pink). 
The interface between the circularly rotating dense disk and 
radially infalling flow (blue region) moves inward from 
$\varpi \simeq 30$ at $t \sim 24000$ to $\varpi \simeq 3$ at $t \sim 26000$.
The radius of the interface oscillates around $\varpi \simeq 10$. 
The infall of dense matter from the inner region of the torus 
($\varpi \simeq 20$) takes place intermittently at $t \sim 26000$ 
and $t \sim 29000$. 
The interval of the infall is nearly equal to the rotation period  
of the dense torus ($t_0 \sim 3079$). 
The intermittency arises from the oscillation of the torus excited by 
the angular momentum exchange between the infalling matter and the torus 
material; the dense torus is kicked by the infalling blob due to the 
angular momentum conservation. 
Note that in figure 7 the location of the density maximum of the torus 
is pushed outward following the infall of dense blobs. 
Short time scale oscillations in the inner region are overlaid to 
this global oscillation. 
%
%
Figure 8 shows the time development of the azimuthally averaged 
radial velocity relative to the mean radial velocity approximated 
as $\langle v_{\varpi} \rangle = - \max{(3 \varpi^{-3}, 0.003)}$ 
at equatorial plane.
Bright region and dark region show the region where 
$\Delta v_{\varpi} = v_{\varpi} - \langle v_{\varpi} \rangle > 0$ 
and $\Delta v_{\varpi} < 0$, respectively. 
The innermost region ($2 \leq \varpi \leq 10$) shows 
quasi-periodic oscillations 
with frequency $f \sim \kappa_{\rm max} / 2 \pi \sim 
1140 (M/M_{\odot})^{-1} 
 {\rm [Hz]}$ where $\kappa_{\rm max}$ 
is the maximum of epicyclic frequency as already pointed out 
by one-dimensional simulations 
(e.g., Matsumoto, Kato \& Honma 1988; 
Kato, Fukue \& Mineshige 1998). 
Significant inflow occurs around $t=25000$, $27000$, $28000$, 
$29500$ and $31000$.

Figure 9 shows the position of the Lagrange test particles uniformly  
distributed in the equatorial plane 
inside $\varpi < 18$ at $t=30319$. 
Spiral arms which appear in figure 9 correspond to the dense arms in the 
density distribution (see figure 3). 
The diamond shaped symbols trace the same test particles. 
The motion of the particles show the development of non-axisymmetric MRI. 
The particle $\sharp$4, for example, has lower angular momentum than 
other particles and spirally infalls. 
The infalling matter deforms the magnetic fields into the BSS shape. 
We can also see the train of spirally 
infalling particles around $x=3, y=0$ at $t=30550$ and $t=30630$. 
Figure 8 indicates 
that accretion to the black hole is not time steady but intermittent. 
%
Figure 8 shows that 
evacuated region is created around $x = 5, y = -5$ at $t=30590$. 
The sharp ridge of the density distribution around $x = 0, y = -7$ in 
figure 3b indicates the interaction between the outgoing wave and the 
infalling matter. In order to show the propagation of outgoing wave, 
we show in figure 10 the distribution of density. 
At $t=30570$, dense region (colored in orange) appears and 
propagates outward. 
As denoted by arrows, density enhancement develops in the evacuated region 
and propagates in the direction opposite to the rotation of the disk. 
In the next subsection, we would like to show that the motions of 
these density enhancements are due to magnetic reconnection.

\subsection{Magnetic Reconnection in the Plunging Region}
%
Figure 11 shows the time evolution of the current density distribution and 
magnetic field lines in the innermost region ($ -10 \leq x \leq 10$, 
$ -5 \leq y \leq 5$). 
In the outer region where magnetic field lines show 
turbulent structure (figure 3b),  
isocontours of current density have fractal like distribution 
(Kawaguchi et al. 2000). 
On the other hand, long current sheets are formed in the inner 
infalling region ($\varpi \leq 8$) where magnetic field lines show 
BSS shape (figure 3d). 
Magnetic energy is accumulated in these current sheets. 
Red regions in figure 11 are active regions 
where the current density is high. 
Active region "A"  is a narrow peak in the current distribution 
(current sheet). 
White arrows denoted by "A" trace the motion of this current sheet. 
Active region "B" has a cusp in the current distribution. 
White arrows denoted by "B" indicate the tip of the reconnected 
magnetic field lines behind the reconnection point. 
At $t = 30590$, the current layer in active region 
"B" coincides with the density 
enhancement shown in figure 10b. 
The lower panels in figure 11 show that 
the cusp-shaped current layer "B" propagates in the direction relatively 
opposite to the direction of disk rotation and that 
current dissipation takes place.
These are the magnetic flaring sites where magnetic energy is released 
by magnetic reconnection. 
The double peak in the current distribution which splits from the cusp 
is a characteristic feature of slow shocks associated with 
magnetic reconnection (see Priest 1982). 

%
Figure 12 shows the time evolution of magnetic field lines. 
Closed magnetic loops (magnetic islands) formed around the cusp of 
magnetic field lines indicate that magnetic reconnection is 
taking place in the current sheet. 
Figure 12 indicates that magnetic reconnection in an accretion disk 
has two types; (a) magnetic reconnection taking place 
inside an elongating magnetic loop and  
(b) magnetic reconnection driven by the interaction of two magnetic loops. 
%
In figure 13, we schematically show the magnetic reconnections in 
solar corona and in accretion disks. 
Arcade type solar flares (e.g., Tsuneta et al. 1992)  
which accompany loop eruption 
belong to the loop elongation type. 
Solar microflares driven by loop interaction produces 
X-ray brightning and X-ray jets (see e.g., Shibata et al. 1992). 
The loop elongation type reconnection can be seen in active region 
"A" at $t=30610$ in figure 11 and figure 12c. Backward propagating
reconnection jet appears as indicated by arrows in figure 10c. 
The active region "B" in figure 11 subjects to both type of reconnection. 
At $t=30590$, the active region "B" collides with the preceding 
magnetic loop (figure 12b) and produces intense current layers. 
Magnetic reconnection taking place in this current layer 
released more energy than active region "A" 
because the current layer "B" has larger volume. 
Although loop interaction releases less energy than arcade eruptions 
in solar corona, loop interaction can release more energy in 
accretion disks because rotational energy of the disk is 
converted to the magnetic energy by such interactions. 
%
Figure 14 shows the time evolution of temperature. 
Magnetic energy is  deposited into the thermal energy and 
heats the plasma around the reconnection site. 
Filamentary shaped hot region which coincide with current sheets 
trace the slow shocks associated with magnetic reconnection. 

%
Figure 15a shows the time evolution of the volume integrated Joule 
heating rate $E_j = \int \eta j^2 dV$ in $ 2 \leq \varpi \leq 6$ and 
$ 0 \leq z \leq 10$. Figure 15b shows the time evolution of 
$\int j^2 dV$. Figure 15c shows the magnetic energy 
$E_m = \int (B^2/ 8 \pi) dV$. 
Figure 15d shows accretion rate at $\varpi = 2.5$. 
Arrows indicate $t = 30590$ and $t = 30630$. 
The largest flare occurs around this time interval. 
Since we turn on  anomalous resistivity 
only when $j/\rho > (j/\rho)_c$, 
magnetic energy is released 
after the accretion rate decreases and the innermost region is 
rarefied.
The Joule dissipation rate 
$\eta j^2$ tends to be zero when the matter 
density of the BSS channel is high. 
This is the reason why $E_j $ has maximum after the peak of 
$\int j^2 dV$. 
Magnetic energy decreases between $t=30590$ and $t=30630$ as 
expected from the magnetic reconnection model. 
Only a small fraction ($ < 1$\%) of the released magnetic energy 
goes into the Joule heating because $\eta \le 10^{-4}$ in this simulation. 
The released magnetic energy is converted to the thermal energy by 
slow shock (see figure 14). 
The volume integrated current 
density correlates well with the mass accretion rate. This is because 
magnetic fields are stretched and amplified by mass accretion. 
In other words, 
parts of the gravitational energy of the accreting matter are stored in the 
current sheet and released by magnetic reconnection.

%
Next, let us compare the X-ray light curve obtained by numerical simulations 
with observations of X-ray shots. 
Figure 16 shows the time variation of the X-ray luminosity computed by 
the optically thin bremsstrahlung luminosity 
$F_x = \int \rho^2 T^{1/2} f(T) dV $ (dashed curve) and Joule heating rate  
$E_j = \int \eta j^{2} dV$ (solid curve) where the integration is 
over the innermost region ($2 \leq \varpi \leq 6, 0 \leq z \leq 10$). 
We introduce cutoff function $f(T)$ where $f(T) = 1$ when 
$0.01 < T < 0.1$ and otherwise $f(T) = 0$. 
The vertical scale of figure 16 is 
arbitrary because $F_x$ and $E_j$ depend on the unit 
density $\rho_b$ and the electric resistivity $\eta_0$ 
(see equation (5)). 
As the matter in dense spiral arms accrete to the innermost region, 
X-ray luminosity gradually increases and has a peak around $t \sim 30100$. 
The time scale of the 
increase of X-ray luminosity is the accretion time scale of dense blobs 
and typically $\sim 10^3 r_g/c \sim 0.01 M/M_{\odot}$ sec 
in our simulation (see figure 7). 
As the dense blob is swallowed into the black hole, 
X-ray luminosity decreases. 
Subsequently, magnetic reconnection takes place in the rarefied region and 
releases the magnetic energy. The soft X-ray excess precedes the 
magnetic reconnection about $500 r_g/c \sim 5 M/M_{\odot} {\rm ms}$.

\section{Summary and Discussion}

We have shown that in the innermost plunging region of black hole accretion 
disks, growth of non-axisymmetric MRI creates bisymmetric spiral (BSS) 
magnetic fields. 
Mass accretion proceeds along these spiral channels. 
As a result, density distribution tends to show one armed spiral structure. 

A current sheet is created inside the BSS channel because the magnetic 
field changes its direction inside the channel. 
The current sheet is built up by converting the gravitational energy of the 
accreting gas into magnetic energy. In other words, magnetic energy is 
accumulated by mass accretion. 
When the BSS channel is rarefied after dense blobs infall, 
magnetic reconnection takes place 
and converts the magnetic energy into heats. 
This mechanism is analogous to the accumulation of magnetic energy by mass 
motion preceding solar flares and protostellar flares 
(e.g., Hayashi et al. 1996). 
Magnetic reconnection can also take place by interaction of two long 
current sheets. 
These X-ray flares in the plunging region can be the origin of large 
amplitude X-ray time variabilities characteristic of black hole candidates. 

Let us compare our numerical results with X-ray observations of Cyg X-1
(Negoro et al. 1995). In X-ray hard states, X-ray flux from Cyg X-1 shows 
largest shots whose interval is typically several seconds 
(Negoro \& Mineshige 2002). 
In our simulation, the largest shot occurred at $0.3 M/M_{\odot}$ 
sec after the initial torus is set up at $\varpi \sim 50$. 
The magnetic reconnection produces outgoing waves and 
cusp-shaped current sheets. 
We have shown that the mass accretion from the circularly rotating 
dense torus is not time steady but intermittent. 
The interval of mass feeding from the torus is the rotation period 
of the torus (figure 7a). 
In our simulation, this time scale corresponds to 
$t_0 \sim 3000 r_g/c \sim 0.03 M/M_{\odot}$ sec. 
If the torus locates at $\varpi_b^{\prime} = 500 r_g$, this interval 
$t_0^{\prime} \sim 0.9 M/M_{\odot}$ sec approaches the interval of 
observed shots if the mass of the black hole is $10 M_{\odot}$. 
Once dense blobs accrete to the innermost region, 
the profile of the shot is essentially determined by the local processes 
in the innermost region. 
Thus the profile of one shot may not depend on the location of the 
center of the torus. 
However, the interval of the shot depends on $\varpi_b$. 
Smaller X-ray shots and fluctuations 
can be created by magnetic reconnections ubiquitous 
in accretion disks. 
Observations of Cyg X-1 indicate that the 
averaged shot profile is time symmetric (Negoro et al. 2001) and 
X-rays become hard at the peak of the shot. 
%
Figure 17 schematically shows the mechanism of X-ray shots. 
The upper left panel shows the accretion stage when dense blobs fall 
into the innermost region. As the density in the innermost region increases, 
the soft X-ray luminosity increases. Current sheets are formed 
inside the BSS channel and magnetic energy is accumulated. 
After the blobs are swallowed into the black hole, the current sheet 
in the BSS channel is rarefied. The soft X-ray luminosity begins to decrease 
because X-ray emitting gas is depleted. 
Largest magnetic reconnection takes place in the rarefied current sheet 
by loop elongation or by loop interaction. 
Since the released magnetic energy is converted to heat, 
X-ray luminosity will suddenly increase and decays
 exponentially. 
In order to compute the light curve after magnetic reconnection, 
we will have to include the effects of heat conduction along the 
magnetic loop which may evaporate the disk material 
(see the numerical simulation of solar flares by Yokoyama \& Shibata 2001). 
We would like to report the results of such simulations in future. 

The BSS magnetic fields in the innermost region of black hole accretion 
disks contribute to the angular momentum transport of the disk 
material. In this region, angular momentum transport due to this 
ordered spiral magnetic fields is more efficient than the angular momentum 
transport by turbulent magnetic fields. 

The dynamical effects of magnetic fields in the plunging region of 
black hole accretion flows have been investigated analytically by 
Krolik (1999), Gammie (1999), Agol \& Krolik (1998, 2000) and simulated by 
Hawley \& Krolik (2001, 2002) and Krolik \& Hawley (2002). 
Our numerical results support their conclusion that the ratio of 
stress to pressure ($\alpha$) has systematic gradient with radius and 
has larger values well inside the plunging region. 

In the innermost region of the disk, the accretion flow 
is dominated by radial advection. 
The radial structure of the innermost region ($\varpi \leq 10 r_g$) 
of accretion flow obtained by three-dimensional MHD simulations 
approaches the global transonic solution of optically thin disks 
with the viscous parameter 
$\alpha \sim 0.1$ (e.g., Narayan et al. 1997). 
The direct numerical simulations such as the one we presented in this 
paper have the potential to do much more than seek agreement with 
$\alpha$-models. 
They can predict the time varitation of black hole accretion flows 
without assuming the phenomenological $\alpha$-parameter. 
In order to confirm the applicability of numerically obtained accretion 
flows to black hole candidates, 
it is essential to compute the X-ray spectrum from numerical results 
and compare them with observations.  
We would like to report the results of such analysis in 
subsequent papers. 

\acknowledgments
The authors thank Drs. S. Mineshige, K. Shibata, H. Negoro, 
T. Sano, and R. Narayan for discussion. 
The authors thank the Yukawa Institute for Theoretical Physics at 
Kyoto University. 
Discussions during the YITP workshop YITP-W-01-17 on 
"Black Holes, Gravitational Lens, and Gamma-Ray Bursts" were useful 
to complete this work. 
Numerical computations were carried out by VPP5000 at NAOJ. 
This work was supported in part by 
Research Fellowships of the Japan Society for the Promotion of Science 
for Young Scientists (13011092, MM), 
and by ACT-JST (RM) of Japan Science and Technology corporation.

\clearpage

\begin{figure}
\epsscale{1.0}
\figcaption[figure1.ps]{
(a) Initial model of the torus. 
(b) The isosurface of density ($\rho = 0.4$) at $t=30590$. 
Numbers show the ($x, y, z$) coordinate. 
\label{Fig.1}}
\end{figure}

\begin{figure}
\epsscale{1.0}
\figcaption[figure2.ps]{
(a) Azimuthally averaged density distribution $\log{\rho}$ (color scale) and 
the poloidal momentum vectors (arrows) at $t=30590$. 
Long arrows are plotted in dark color. 
(b) Azimuthally averaged density distribution
at $t=0$, $t=21502$ and $t=30590$. 
\label{Fig.2}}
\end{figure}

\begin{figure}
\epsscale{1.0}
\figcaption[figure3.ps]{
The equatorial density distribution (color scale) and 
magnetic field lines projected onto the equatorial plane 
(grey curves) at $t=30590$.
(a), (b) Global structure inside $60 r_g$.
(c), (d) Inner region inside $10 r_g$. 
Color scale of logarithmic density $\log{\rho}$ is shown at the bottom of 
each panel. 
\label{Fig.3}}
\end{figure}

\begin{figure}
\epsscale{1.0}
\figcaption[figure4.ps]{
Three-dimensional magnetic field structure at $t=30590$. 
White curves show magnetic field lines, blue surface show the 
isosurface of the density ($\rho = 0.4$), and yellow plane shows the 
contour of the equatorial density. 
The size of the box is $-10 \leq x \leq 10$, $-10 \leq y \leq 10$ 
and $-10 \leq z \leq 10$. 
\label{Fig.4}}
\end{figure}

\begin{figure}
\epsscale{0.5}
\figcaption[figure5.ps]{(a) The time developments of the volume averaged 
plasma $\beta \equiv \langle P \rangle / \langle B^2 /8 \pi \rangle  $. 
(b) The time developments of the ratio of the volume averaged 
Maxwell stress to pressure, 
$ \alpha_B \equiv - \langle B_{\varpi} B_{\varphi}/ 4 \pi \rangle 
/ \langle P \rangle$.
(c) The time developments of the accretion rate at $ \varpi = 2.5 r_g$.  
The solid curves show the average in the inner region 
$ 4 \leq \varpi \leq 10$ and $0 \leq z \leq 1$. 
The dotted curves show the average in the outer region 
of the disk, $ 20 \leq \varpi \leq 40$, and $0 \leq z \leq 3$. 
\label{Fig.5}}
\end{figure}

\begin{figure} 
\epsscale{1.0}
\figcaption[figure6.ps]{
Radial profiles of density, gas pressure, magnetic pressure, 
radial velocity and specific angular momentum 
averaged in azimuthal direction,  
vertical direction ($0 < z < 0.3$) and in time. 
Solid curves show quantities averaged in $29000 < t < 31000$ 
and dashed curves show quantities averaged in $28000 < t < 29000$. 
The dotted curve in (d) shows the Keplerian angular momentum 
distribution. 
\label{Fig.6}}
\end{figure}

\begin{figure}
\epsscale{0.9}
\figcaption[figure7.ps]{
Time development of the azimuthally averaged density 
at the equatorial plane. 
The range of the color scale is $-0.5$ (blue) 
$\leq \log{\rho} \leq 0$ (pink). 
\label{Fig.7}}
\end{figure}

\begin{figure}
\epsscale{0.9}
\figcaption[figure8.ps]{
Time development of the azimuthally averaged radial velocity 
at the equatorial plane $\Delta v_{\varpi} = v_{\varpi} - 
\langle v_{\varpi} \rangle$ where 
$\langle v_{\varpi} \rangle = - \max{(3 \varpi^{-3}, 0.003)}$.
The range of the gray scale is 
$-0.07$ (dark) $\leq \Delta v_{\varpi} \leq 0.07$ (bright).
\label{Fig.8}}
\end{figure}

\begin{figure}
\epsscale{0.8}
\figcaption[figure9.ps]{
Positions of Lagrange test particles on the equatorial plane. 
Large symbols ($\sharp 1 \sim \sharp 6$) trace the same particles. 
\label{Fig.9}}
\end{figure}

\begin{figure}
\epsscale{1.0}
\figcaption[figure10.ps]{
Density distribution in the region $ 2 \le \varpi \le 10$. 
Color scale in each panel shows the scale of $\log{\rho}$. 
Numbers at the upper right corner of each panel show time. 
Arrows indicate the reconnection jet. 
\label{Fig.10}}
\end{figure}

\begin{figure}
\epsscale{0.5}
\figcaption[figure11.ps]{
Time evolution of the current density (color scale) 
and magnetic field lines (white curves). 
Red area show active regions where the current density is high. 
White arrows "A" trace the motion of a current sheet. 
White arrows "B" indicate the tip of the reconnected magnetic field lines. 
The size of each box is $-10 \leq x \leq 10$ and 
$-5 \leq y \leq 5$. 
Numbers show time. 
\label{Fig.11}}
\end{figure}

\begin{figure}
\epsscale{1.0}
\figcaption[figure12.ps]{
Magnetic field lines projected onto the equatorial plane. 
Numbers show time. 
The size of the box is $-5 < x < 5$ and $-5 < y < 5$. 
\label{Fig.12}}
\end{figure}

\begin{figure}
\epsscale{1.0}
\figcaption[figure13.ps]{
A schematic picture of magnetic reconnection in solar flares and 
in accretion disks. 
\label{Fig.13}}
\end{figure}

\begin{figure}
\epsscale{1.0}
\figcaption[figure14.ps]{
Distribution of logarithmic equatorial temperature. 
The size of each box is $2 \leq \varpi \leq 10$. 
\label{Fig.14}}
\end{figure}

\begin{figure}
\epsscale{1.0}
\figcaption[figure15.ps]{Time evolution of (a) the volume integrated 
Joule heating rate, 
(b) the  volume integrated current density,  
(c) the magnetic energy, and 
(d) the accretion rate at the $\varpi = 2.5$. 
Arrows indicate $t = 30590$ and $t = 30630$. 
\label{Fig.15}}
\end{figure}

\begin{figure}
\epsscale{1.0}
\figcaption[figure16.ps]{
Time variation of the X-ray luminosity computed by the simulation. 
Solid curve shows the Joule heating rate. Dashed curve shows the 
soft X-ray luminosity by bremsstrahlung. 
\label{Fig.16}}
\end{figure}

\begin{figure}
\epsscale{1.0}
\figcaption[figure17.ps]{
A schematic picture of the mechanism of X-ray shots. 
Soft X-ray excess precedes the hard X-ray flare. 
\label{Fig.17}}
\end{figure}

\end{document}